\documentclass[journal,draftclsnofoot,onecolumn,12pt,twoside]{IEEEtran}
\usepackage{mathrsfs}
\usepackage{amsfonts}
\usepackage{graphicx,cite,epsfig,amssymb,amsmath}
\usepackage{color,xcolor}
\usepackage{pifont}
\usepackage{stmaryrd}
\usepackage{setspace}
\usepackage{subfigure}
\usepackage{cite}
\usepackage{array}\usepackage{float}
\usepackage{multirow}
\usepackage{algorithmic,algorithm}
\usepackage{subeqnarray}
\usepackage{cases}
\usepackage{url}
\usepackage{graphicx}
\usepackage{epstopdf}
\usepackage{colortbl}
\usepackage{background}
\SetBgContents{}
\providecommand{\tabularnewline}{\\}
\floatstyle{ruled}
\newfloat{algorithm}{tbp}{loa}
\providecommand{\algorithmname}{Algorithm}
\floatname{algorithm}{\protect\algorithmname}

\newcommand{\tabincell}[2]{\begin{tabular}{@{}#1@{}}#2\end{tabular}}

\makeatletter
\definecolor{mygray}{gray}{0.7}
\providecommand{\tabularnewline}{\\}
\floatstyle{ruled}
\newfloat{algorithm}{tbp}{loa}
\providecommand{\algorithmname}{Algorithm}
\floatname{algorithm}{\protect\algorithmname}

\makeatother
\begin{document}

\title{Embedding LTE-U within Wi-Fi Bands for Spectrum Efficiency Improvement}

\author{\small{Qimei Chen}$^{1}$, Guanding Yu$^{1}$, Hesham M. Elmaghraby$^{2}$,  Jyri H$\ddot{a}$m$\ddot{a}$l$\ddot{a}$inen$^{3}$, and Zhi Ding$^{2}$\\
1. College of Information Science and Electronic Engineering, Zhejiang University, Hangzhou 310027, China\\
 2. Department of Electrical and Computer Engineering, University of California, Davis, CA 95616, USA \\
 3. School of Electrical Engineering, Aalto University, Aalto 00076, Finland
 }
\maketitle
\vspace{-2cm}
\begin{abstract}
Driven by growing spectrum shortage,  Long-term Evolution in unlicensed spectrum (LTE-U)
has recently been proposed as a new paradigm to deliver better performance and experience
for mobile users by extending the LTE protocol to unlicensed spectrum.
In the paper, we first present a comprehensive overview of the LTE-U technology, and discuss the practical challenges it faces.  We summarize the existing LTE-U operation modes and
analyze several means for LTE-U coexistence with Wi-Fi medium access control protocols.
We further propose a novel hyper access-point (HAP) that
integrates the functionalities of LTE small cell base station and commercial Wi-Fi AP for deployment by
cellular network operators.  Our proposed LTE-U access embedding within the Wi-Fi protocol is non-disruptive
to unlicensed Wi-Fi nodes and demonstrates performance benefits as a seamless and novel LTE and Wi-Fi coexistence technology in unlicensed
band. We provide results to demonstrate the performances advantage of this novel LTE-U proposal.
\end{abstract}

\section{Introduction}

The rapid growth of mobile wireless applications and consumers
continues to strain the limited cellular network capacity and has motivated the exploration of next generation 5G wireless networks.
The mobile industry predicts a $1000$x data
traffic growth by 2020. To meet this anticipated data growth demand, both industry and academia are exploring various
advanced solutions to boosting network capacity while continually providing high-level user experience to wireless customers.
One natural direction is developing new technologies to improve the efficiency of limited licensed spectrum.
Such popular advances include small cell base station (SBS) coverage, massive multiple-input multiple-output
links, and device-to-device communications.
Despite these exciting solutions, one key  obstacle to network capacity expansion still lies in
 the scarce licensed spectral resources.

To overcome spectrum shortage, LTE in unlicensed spectrum (LTE-U) technology has been proposed and is currently under
consideration by the 3GPP into its future standards.
LTE-U technology allows users to access both licensed and unlicensed spectra under a unified LTE network infrastructure.
It can provide better link performance, medium access control (MAC), mobility management, and larger coverage than
simple Wi-Fi offloading \cite{traffic,offload}.

Despite the many advantages of LTE-U, it still faces certain technical challenges. One primary issue is the coexistence with the incumbent Wi-Fi systems.
On the one hand, LTE systems are specifically designed to operate in the licensed spectrum under the centralized control of
network units based on non-contention MAC protocols
to prevent packet collision among subscribers.
On the other hand,  Wi-Fi users rely on carrier sensing multiple access with collision avoidance (CSMA/CA) to reduce packet collision and
use a contention based MAC protocol, namely distributed coordination function (DCF),
to resolve package collision through a random backoff mechanism.
Therefore, how to ensure a fair and harmonious coexistence environment for both networks becomes a
major challenge for LTE-U.

First, for the two conflicting MAC technologies to coexist in the same unlicensed band, their mutual
impact and interference must be carefully controlled and managed.
It is important that a) Wi-Fi transmissions
will not frequently collide with LTE-U transmissions; b) LTE-U transmissions
will not lead to a substantial drop of Wi-Fi throughput.
Secondly, we should require a coexistence proposal of LTE-U and Wi-Fi so as not to demand
fundamental changes to the existing protocols defined in LTE and Wi-Fi standards.
This requirement is essential to avoid any impractical retrofitting of IEEE 802.11-compliant
Wi-Fi radios, or the re-design of existing 3GPP standards.

To overcome such challenges, we will first propose a new wireless access point (AP) architecture, namely  hyper-AP (HAP), which is deployed by cellular operators and functions simultaneously as an LTE SBS and
a Wi-Fi AP.  By integrating both LTE and Wi-Fi functions within the same HAP node, it can then jointly
coordinate the spectrum allocation and the interference management for Wi-Fi and LTE-U.
Based on the use of HAP, we develop a novel LTE-U and Wi-Fi coexistence mechanism by
embedding LTE-U signalling within the existing Wi-Fi protocol for seamless integration. Specifically,
the HAP will provide LTE user access in the contention free period (CFP) under a centralized coordinator,
whereas the HAP continues to provide normal Wi-Fi user access through its contention period
(CP) using the traditional CSMA/CA mechanism.
The main merit of the newly proposed HAP is that it can well support the standalone mode for LTE-U transmission,
which is found to be a formidable hurdle in many other coexistence mechanisms.

In what follows, we shall first provide an LTE-U technology overview, including operation modes,
coexistence technologies for LTE and Wi-Fi, and current considerations of combining LTE-U technology within
Wi-Fi MAC protocols. Next,  we introduce the new wireless HAP that provides both LTE-U access
and Wi-Fi access in shared unlicensed band.
To improve the system performance, our key contribution is the novel LTE and Wi-Fi coexistence mechanism that embeds LTE-U channel access
within a standardized Wi-Fi protocol.
We shall also provide simulation results to
illustrate the achievable network performance gain in terms of Wi-Fi per-user throughput and overall system throughput.
Compared with networks under traditional coexistence technology without LTE-U deployments,
the proposed network access demonstrates better performance for both LTE-U and Wi-Fi user groups.

\section{Overview of LTE-U Technology}

\subsection{LTE-U Operation Modes}
Presently, there are mainly three operation modes proposed for LTE-U networking
based on different control channel configurations and supplementary links:
supplementary downlink (SDL),  carrier aggregation (CA) time division LTE (TD-LTE), and  standalone LTE-U \cite{coe1},\cite{coexistence}.

In SDL mode, the licensed spectrum serves as the anchor carrier.
When needed, the unlicensed spectrum is used for additional downlink data transmission since the downlink
demand is often higher than uplink.
Since the SDL mode merely utilizes the unlicensed spectrum as a downlink supplement, carrier aggregation technique can be applied
with few changes on the existing LTE standards. 

In carrier aggregation TD-LTE mode, the unlicensed spectrum is used as an auxiliary TDD channel to carry both downlink and uplink traffic.
Its control channels remain on the licensed spectrum.
Once again,  CA techniques can be adopted to execute the carrier aggregation TD-LTE mode.

In standalone LTE-U mode, both data and control signals from LTE-U nodes
are transmitted on unlicensed spectrum without relying on the additional licensed spectrum.
This mode can be used in the areas without guaranteed coverage of licensed cellular infrastructure,
and hence is a more flexible form of LTE-U.
Moreover, this mode also lacks an intelligent and centralized control signalling on licensed spectrum to ensure
reliable transmission, resource utilization, channel measurement, among others.


\subsection{LTE and Wi-Fi Coexistence Technologies} \label{section_coexistence}
Wi-Fi and LTE have substantially different PHY/MAC protocols, making their joint coordination
exceptionally challenging. In a nutshell, the LTE system uses a centralized MAC protocol
which allocates a non-overlapping set of physical resource blocks in time-frequency domain to mobile users within a cell.
There is no contention among users within such centralized multiple access in both downlink and uplink.
On the other hand, the Wi-Fi MAC protocol uses a decentralized, contention-based, random access mechanism
based on CSMA/CA and DCF.
If properly designed, LTE-U networks should be a good spectrum-sharing partner with the
incumbent Wi-Fi networks within the unlicensed band \cite{neighbor}.
Next, we will introduce several known LTE and Wi-Fi coexistence mechanisms, which are also summarized in Table \ref{tab_coe}.

\subsubsection{Channel Splitting Coexistence}
In some cases, the unlicensed spectrum is sufficiently wide.
It becomes possible for the LTE-U SBS to choose the cleanest channel, i.e., the channel exhibiting the lowest sensed interference power level,
based on power and activity measurements.    
In this case, the SBS is almost free of interference from its
neighboring Wi-Fi devices if the Wi-Fi APs are given a ``channel-clearing'' request by the SBS.
Qualcomm has developed a simple and efficient dynamic channel selection method \cite{neighbor}.
Other works have also discussed ways to leverage existing underlay techniques in both LTE-U and Wi-Fi systems to ensure the least congested channel selection.
We shall note that, although the unlicensed spectrum is not unlimited, there is still a chance that it has not been fully utilized by Wi-Fi users.
There may exist little Wi-Fi activity in some locations such as certain outdoor environments.
Therefore,  dynamic channel selection may  suffice to meet the coexistence requirement in various low or medium-density Wi-Fi
deployment scenarios.

\subsubsection{Channel Sharing Coexistence}
When the Wi-Fi APs are densely deployed, it is very likely that the SBS has to share the unlicensed channel with  Wi-Fi systems.
According to the regional regulatory requirements, co-channel coexistence mechanisms can be separated into two kinds:
listen-before-talk (LBT) and non-LBT markets.

In  LBT markets, such as Europe and Japan, the mechanism of CSMA/CA will be used.
Before data transmission, the cognitive SBS will attempt to access the unlicensed spectrum to assess whether
there exists sufficient access opportunities. When the unlicensed spectrum is deemed idle, the SBS will activate; otherwise,
the SBS will back off one period.
Given the simplicity of this CSMA/CA protocol, it is not effective in highly dense Wi-Fi networks that rarely have
idle channels.

In non-LBT markets, such as the United States, South Korea, China, and India,
several mechanisms have also been proposed to facilitate the coexistence.
Qualcomm has proposed a duty-cycle based carrier sensing adaptive transmission (CSAT) mechanism that
measures channel utilization by neighboring nodes and adopts the on/off duty cycle of secondary cells in the unlicensed band \cite{neighbor}.
The duty cycle period is determined by LTE MAC control elements, and is adaptively chosen based on the number of measured
active Wi-Fi APs and their channel occupancy to allow a fair sharing between LTE-U and Wi-Fi.
In \cite{coe6}, a similar duty cycle mechanism called LTE muting was proposed to silence some LTE-U subframes for
Wi-Fi usage.  The almost blank subframes mechanism in existing LTE standard can also be exploited to
enable the duty cycle mechanism \cite{coe8}.
Furthermore, the authors in \cite{bib_Yun} proposed to utilize multiple
antenna configurations and to exploit different frame structures between the two systems for efficient coexistence.
The cognitive radio schemes have also been taken into account for LTE and Wi-Fi coexistence in \cite{bib_Guan}.

Generally, the proposed mechanisms in non-LBT markets
are more compatible with the existing LTE standards and are simpler to implement than the LBT mechanism.
Moreover, in the high density Wi-Fi environment, the LBT mechanism is likely to fail since open channel opportunities
for LTE-U are slim. In such cases,  non-LBT mechanisms can achieve a good performance by carefully selecting
the duty cycle (or muting) period.

\begin{table}[!hbp] %
\renewcommand{\multirowsetup}{\centering}
\caption{LTE and Wi-Fi coexistence technologies} \label{tab_coe}
\small
\center
\begin{tabular}{|c|c|c|c|}
\hline
\rowcolor{mygray}
 \bf{Types} & \bf{Features} & \bf{Ideas} & \bf{Examples} \\
\hline
 \tabincell{c}{Channel-\\ Splitting} & \tabincell{c}{Channel\\ selection} & \tabincell{l}{SBS selects the channel having the \\ lowest interference power to avoid \\interference with nearby nodes.} & \tabincell{c}{Dynamic frequency\\ selection \cite{neighbor}}\\
\hline
\multirow{5}{2.2cm}{Channel-Sharing}
& LBT & \tabincell{l}{\hspace{1em}SBS adopts the CSMA/CA with\\ \hspace{1em}collision avoidance mechanism\\ \hspace{1em}to access unlicensed spectrum.} & \tabincell{c}{Dynamic duty cycle\\ with LBT \cite{DLBT}} \\
\cline{2-4}
& non-LBT & \tabincell{l}{SBS senses the channel\\ utilization then adopts\\ duty cycle of the unlicensed\\ spectrum.}  & \tabincell{c}{Duty cycle based\\ CSAT \cite{neighbor}; \\ LTE muting \cite{coe6, coe8, bib_Yun, bib_Guan}} \\
\hline
\end{tabular}
\end{table}

\subsection{Coexistence from Wi-Fi Perspective} \label{section_structure_A}

Thus far, our discussions have centered on the necessary aspects of an LTE-U system to
coexist with Wi-Fi networks in the unlicensed band. Here, we will briefly summarize how
Wi-Fi MAC protocols may be affected by the presence of coexisting LTE-U
and discuss their coexistence challenges.

We start with the default and the most commonly utilized Wi-Fi channel access mechanism known as DCF.
Under the DCF mechanism, pure Wi-Fi users will have less opportunities to access the channel if they sense that LTE users are continuously occupying certain channels. Even with LBT or non-LBT mechanisms at the LTE side,
Wi-Fi transmissions could still be disrupted  because of the well-known exposed-node problem, which occurs when a node is prevented from sending packets to other nodes owing to a neighboring transmitter.
Conversely,  LTE transmissions will also be interfered by the Wi-Fi systems under DCF  for the same reason.

Despite the popularity of the DCF mechanism, point coordination function (PCF) is a well-known optional protocol capability for Wi-Fi infrastructure mode.
Under the PCF mechanism, there exists a CFP followed by a CP in each periodic access interval and Wi-Fi nodes can be centrally coordinated by an AP for controlled channel
access with a point coordinator (PC) in the CFP.  Although the PCF option has seldom been used until now,
its centralized, contention-free nature provides a perfect mechanism for LTE-U channel sharing without
interference from contention based Wi-Fi transmission.
By activating PCF, the CFP can be used by LTE-U channel access whereas the CP can be continually occupied by Wi-Fi nodes
in sharing the unlicensed band without mutual interference.

We further note that, in addition to the classic DCF and PCF,  a more versatile hybrid coordination function (HCF) is another option for infrastructure  Wi-Fi
access.  Unlike in PCF, the CFP in the HCF includes several contention
free intervals known as HCF Controlled Channel Access (HCCA) transmission opportunity (TXOP).
Moreover, HCF based Wi-Fi nodes access the Wi-Fi channels by utilizing a QoS-aware hybrid coordinator (HC).
To ensure QoS of LTE-U users,  medium access proportion of each node may be allocated according to the priority of services.
Because of the flexibility and the self-control ability of HCF, it is more suitable to LTE-U applications given
only unlicensed band.

\section{LTE-U Embedded in Wi-Fi Protocols}
In this section, we first propose a novel HAP for LTE and Wi-Fi coexistence and then introduce an innovative LTE and Wi-Fi coexistence mechanism by embedding LTE-U within existing
Wi-Fi protocol.

\subsection{A New LTE-U Framework}


One major obstacle to LTE-U is the lack of centralized coordinator in the unlicensed band, making it difficult
to manage mutual interference between the two channel sharing networks.
To tackle this problem, we propose a novel wireless AP, called HAP, which integrates both the Wi-Fi AP and SBS functionalities
into a single networking unit.
The Wi-Fi side of the HAP exclusively accesses the unlicensed spectrum,
while the LTE side of the HAP can access both licensed and unlicensed bands.
Because the HAP is deployed by the same network service provider,
it can more efficiently mitigate the mutual interference between the two networks
through careful management of both licensed and unlicensed bands.

\begin{figure}
\begin{centering}
\includegraphics[width=0.7\textwidth]{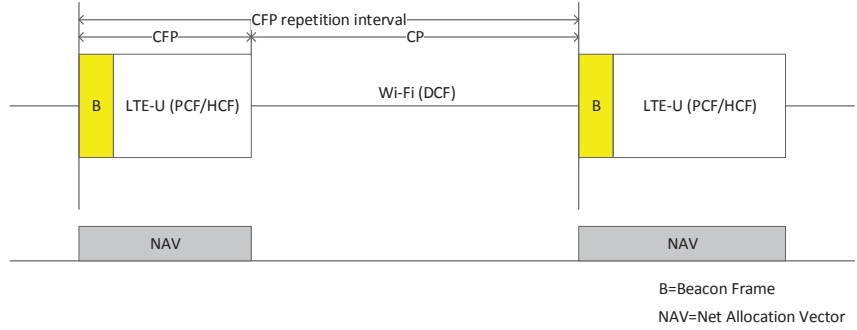}
\par\end{centering}
\caption{LTE-U embedded in Wi-Fi protocol.}\label{fig_unlicensd_model}
\end{figure}

We propose an approach different from the existing LTE-U mechanisms, which is depicted in Fig. \ref{fig_unlicensd_model}.
In the CFP, a centralized coordinator controls the resource allocation scheme. When one LTE-U radio transmits during CFP,
other LTE-U and Wi-Fi radios are muted. In this way, interference between different radios and from the Wi-Fi transmission can be avoided.
On the other hand, since the transmission periods for both LTE and Wi-Fi users are completely separated,
Wi-Fi users will no longer need to compete for unlicensed resources with LTE users.
Hence, the performance of the Wi-Fi network can be separately controlled and managed.

The proposed LTE and Wi-Fi coexistence technique functions as follows. When an HAP is on, it shall first scan the unlicensed band and select the channel with the least background interference power
and the lowest user occupancy.
Thereafter, each HAP determines its CFP and CP allocations for the chosen channel at a CFP repetition interval
according to its traffic needs and other QoS considerations.
Once the setup of channel-sharing between contention-free LTE-U and contention-based
Wi-Fi completes, LTE-U and Wi-Fi users can access the unlicensed spectrum via the CFP and the CP, respectively.

In the following, we will  present two LTE-U embedding  proposals: unlicensed carrier aggregation (UCA) TD-LTE and standalone LTE-U operation  modes.
We note that the SDL mode is a special case of the carrier aggregation TD-LTE
without the uplink transmission in LTE-U.  Therefore, we do not devote special
attention to the SDL mode henceforth. We shall also note that, along with the other LTE-U networks, the following two modes
are better suited to support delay-tolerant services due to the volatility
of the Wi-Fi traffic and the uncertainty of background interference on the unlicensed spectrum.

\subsection{Unlicensed Carrier Aggregation of TD-LTE}

%

\begin{figure}
\begin{centering}
\includegraphics[width=1\textwidth]{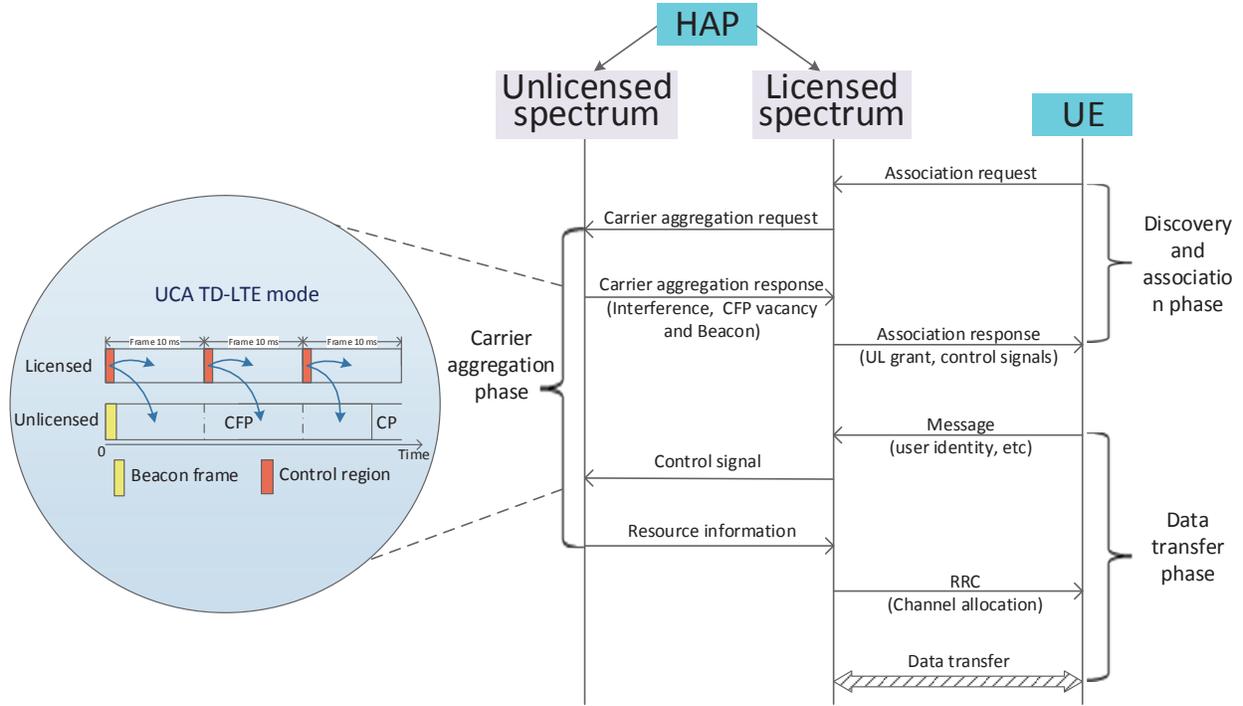}
\par\end{centering}
\caption{Procedure for UCA 
TD-LTE mode.}\label{fig_LTEU-CA}
\end{figure}

We now demonstrate how to implement the UCA TD-LTE mode in the proposed HAP.
As shown in the carrier aggregation phase of Fig. \ref{fig_LTEU-CA}, the HAP will first read its Wi-Fi
beacons, which are located at the start of each CFP. The beacons contain the information of the time stamp and the CFP length.
The HAP then will decide whether and when to aggregate the unlicensed and licensed bands according to the beacon, i.e., based on the available transmission interval calculated by the time stamp and CFP length embedded in the beacon.
The decision is then signalled to the LTE-U users in the licensed downlink control channel (DCCH).
Since the unlicensed spectrum is only used for
data payload transmission (i.e., to expand the downlink shared channel (DSCH) bandwidth),
both PCF and HCF mechanisms can be adopted.

The signalling exchange procedure of the proposed scheme is also shown in Fig. \ref{fig_LTEU-CA}.
First, an association request will be sent by a user equipment (UE). After receiving the association request, the
HAP will search for channels with the lowest interference power and the highest CFP vacancy, then
decide whether and when to aggregate these unlicensed bands as LTE-U channels based on its time-stamp and CFP length,
and later respond to the user with its control signals and an uplink (UL) grant.
Thereafter, a signal carrying the user identity will be sent to the HAP,
and the HAP will send Radio Resource Control (RRC) signal to decide the channel allocation given its licensed bands and the newly acquired carrier aggregation resources.

\subsection{Standalone LTE-U}


We now propose a standalone (SA) LTE-U mode in our HAP framework. In the SA LTE-U mode,
both the control and data channels of the LTE-U system must reside in the unlicensed band without the help of
an auxiliary licensed spectrum.

In this mode, it is likely that the allowable duration of each TXOP
is too short to accommodate a LTE frame (i.e., $10$ ms).   Thus,
to properly embed LTE-U into the Wi-Fi networks,
the active length of each LTE frame should be adapted.
In light of standardized LTE transmissions, we shall exploit the
discontinuous reception (DRX) and discontinuous transmission (DTX) mechanisms in the LTE protocol for LTE-U embedding \cite{DX2}.


According to the existing DTX and DRX standard formats,
the LTE frame can be shortened through silencing to fit into a TXOP duration.
By shortening an LTE frame to fit within a TXOP interval, we are able to embed SA LTE-U transmission into an existing Wi-Fi timing structure in
the unlicensed band. In fact, the Wi-Fi only users can be fully oblivious to the presence of SA LTE-U without the need for any protocol
or standard readjustment.

\begin{figure}
\begin{centering}
\includegraphics[width=1\textwidth]{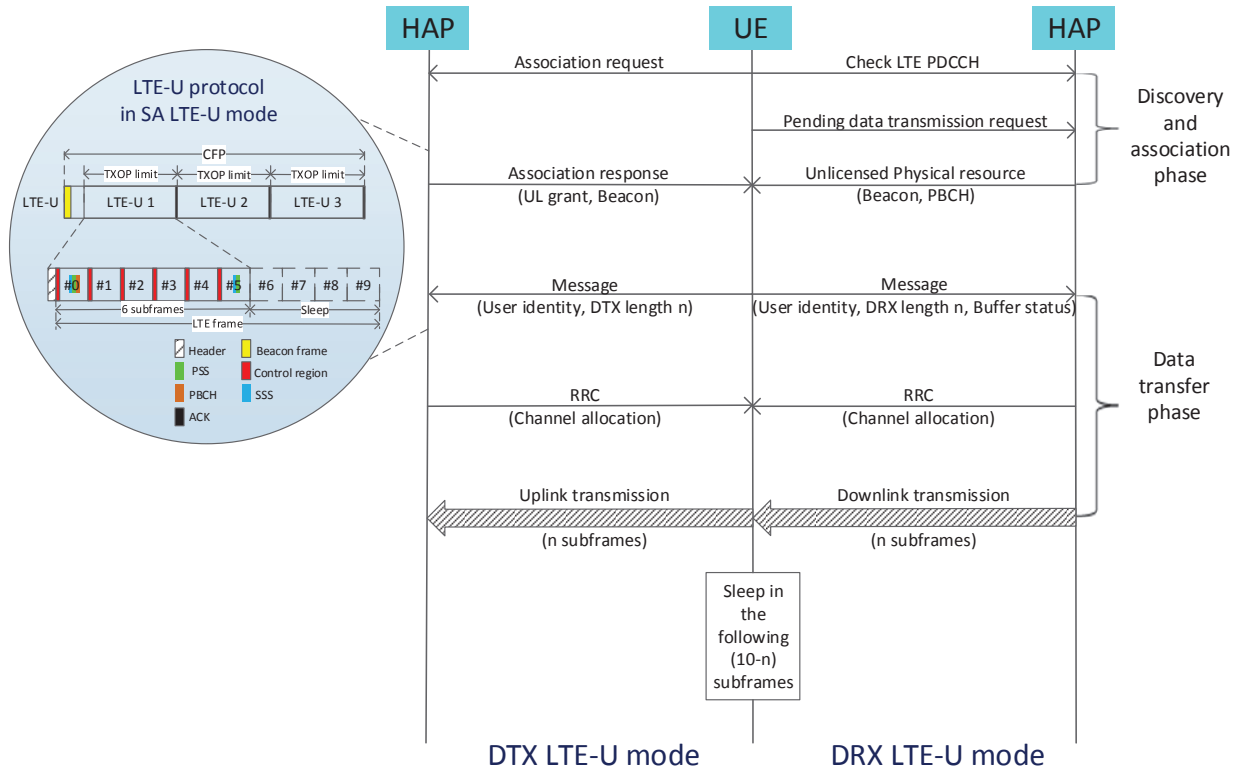}
\par\end{centering}
\caption{Procedure for SA LTE-U mode.}\label{fig_DTX_DRX_LTEU}
\end{figure}

Fig. \ref{fig_DTX_DRX_LTEU} shows an example of a shortened LTE frame with $6$ subframes that are embedded into the TXOP duration. Each LTE  subframe begins with a control region that covers $1$ to $3$ OFDM symbols. The primary and secondary synchronization signals (PSS and SSS, respectively) of the SA LTE-U
are transmitted in subframes $0$ and $5$ and the physical broadcast channel (PBCH) is transmitted in subframe $0$.
Because subframes $0$ and $5$ carry crucial synchronization information, we shall define
each TXOP duration for LTE-U to span at least $6$ subframes\footnote{According to the IEEE 802.11n standard,
TXOP duration is in the range from $32~\mu s$ to $8160~\mu s$ in increments of $32~\mu s$.} or 6 ms.
The beacons in Wi-Fi HCF carry the system information such as the time stamp, the length of CFP,
and the starting time and the maximum length of each TXOP.
Before each (shortened) SA LTE frame, there should also be a header to help with the frame synchronization. Moreover, an ACK signal will be sent at the end of each TXOP.
In what follows, we will show the operation process for both uplink and downlink.

The part of DTX LTE-U mode in Fig. \ref{fig_DTX_DRX_LTEU} shows the process of uplink signal transmission with the DTX mechanism. In the discovery and association phase, a UE will first send an association request to the serving HAP once it has data to transmit. The serving
HAP will respond to the UE with the
beacon information and a UL grant. Thereafter, the UE will check the Wi-Fi Beacon and determine
whether to work with that HAP. After the UE joins LTE-U,
the signal transmission will go into the phase of normal data transfer,
during which both the user identity and DTX length $n$ will be sent to the HAP.
On the unlicensed band, HAP also uses SA LTE-U to exchange RRC information with the UE.
Finally, the uplink and downlink transmissions will stay active for $n$ subframes.
The UE will return to DTX mode in the ensuing $10-n$ subframes, and then wake up
again to send more data.

Similarly, the part of the DRX LTE-U mode in the Fig. \ref{fig_DTX_DRX_LTEU} shows the signalling
exchange procedure of the downlink transmission with the DRX mechanism.
A UE will wake up periodically to check PDCCH. If the UE cannot monitor the PDCCH, it would
return to sleep again; otherwise, the UE would send the pending data transmission request to the serving
HAP. Upon receiving the pending data transmission request, the serving HAP will allocate physical resources on the unlicensed spectrum to the UE.  From the beacon information and the LTE-U broadcast channel,
the UE can locate the LTE-U control channels. Information such as UE identity, buffer status,
and the DRX length $n$ will be sent to
the serving HAP. Accordingly, the HAP determines the resource allocation.
Finally, the shortened $n$ subframes will carry downlink and uplink transmissions, and the UE
will sleep for the next $10-n$ subframes in DRX before waking up for more downlink reception.

\section{Performance Evaluation}

\begin{table}[]
\caption{System Parameters}\label{parameter}

\centering{}%
\small
\begin{tabular}{|c|c|c|c|}
\hline
Parameters & Settings & Parameters & Settings \tabularnewline
\hline
\hline
Noise power & $-174$~dBm/Hz & PHY header & $192$ bits  \tabularnewline
\hline
Path loss exp. & $5$ & MAC header & $224$ bits \tabularnewline
\hline
Transmit power & $30$~dBm & Propagation delay & $20$ $\mu s$ \tabularnewline
\hline
Payload length & $1500$ byte & SIFS & $16$ $\mu s$ \tabularnewline
\hline
Bandwidth & $20$ MHz & DIFS & $50$ $\mu s$  \tabularnewline
\hline
Min. backoff window & 16 & Slot time & $9$ $\mu s$ \tabularnewline
\hline
Max. backoff window & $1024$ & ACK & $112$ bits + PHY header \tabularnewline
\hline
Max. backoff stage & $6$ & RTS & $160$ bits + PHY header \tabularnewline
\hline
WiFi channel bit rate  & $130$ Mbps & CTS & $112$ bits + PHY header \tabularnewline
\hline
\end{tabular}
\end{table}

In this section, we will consider a two-tier network in which HAPs operate under the coverage of a
macrocell. We deploy $N$ Wi-Fi users and $M$ LTE-U users according to Poisson distribution
within the coverage of an HAP. Each user will associate with the proper serving
HAP according to standard LTE association strategies, and the licensed bands are orthogonally allocated to the macrocell as well as each HAP.
We let the Wi-Fi be the most popular IEEE 802.11n protocol in
the $5$ GHz unlicensed band with $20$ MHz bandwidth and certain useful parameters can be found in \cite{ave_LTE}.
We let the HAP coverage have a radius of $100$ m.
The length of the CFP repetitional interval is $100$ ms, which is the default value in the IEEE 802.11n.
The channel fading between LTE-U users and the HAP follows standard 3GPP fading model \cite{fading}. The parameters are listed in Table \ref{parameter}.

In our performance evaluation tests, we compare the HAP based network performance against two
benchmark access schemes.  Specifically, our proposed LTE-U network based on the HAP framework is
compared against  the pure Wi-Fi access without LTE-U deployment (labelled as ``original'' henceforth in all figures).
The LTE-U access based on the LBT protocol is also included.
For simplicity, we do not differentiate specific traffic types for each LTE user and let available channel resources be equally allocated among LTE-U and Wi-Fi users. In other words, we let the fractions of CFP and CP in the HAP network are $\frac{M}{M+N}$ and $\frac{N}{M+N}$, respectively. Note that this assignment is simple but not necessarily optimum.





Fig. 4(a) presents the comparative results of average Wi-Fi per-user throughput under
the three network service protocols.
Recall that the ``original'' network denotes the pure Wi-Fi case without serving LTE-U users,
which naturally delivers the highest Wi-Fi throughput among all three scenarios.
Because of the channel contention, the per-user Wi-Fi throughput decreases as the number of Wi-Fi users grows
under each of the three protocols.
What is more interesting is that, for smaller number of LTE-U users ($M$ = 5), the proposed HAP can improve the per-user Wi-Fi throughput by 20\%- 30\%
above that of the LBT protocol. For larger number of LTE-U users ($M = 10$), the improvement becomes more significant (30\% - 70\%).
The reason is that LBT is still a contention-based protocol whose performance would deteriorate as the number of competing users grows. On the
other hand, our newly proposed HAP-based access protocol divides the LTE-U spectrum access orthogonally from the Wi-Fi user access such that
the cross-network interference between LTE-U and Wi-Fi is avoided. The results in Fig. 4(a) demonstrate that the proposed HAP coverage
can better preserve Wi-Fi network throughput than the LBT scheme.

\begin{figure}
\center
\subfigure[Per-user Wi-Fi throughput under different networks.]{

\begin{minipage}[b]{0.5\textwidth}

\includegraphics[width=10cm]{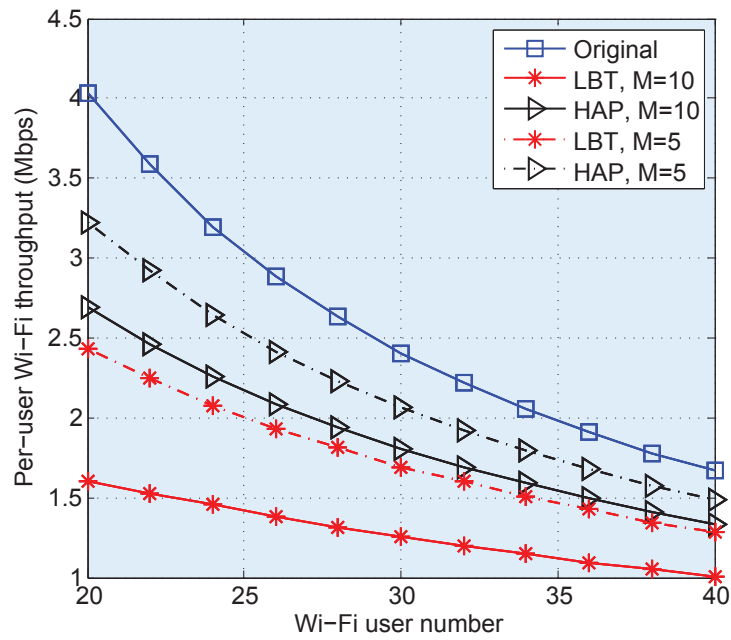}

\end{minipage}

}

\subfigure[LTE-U, Wi-Fi, and total throughput comparison in unlicensed spectrum under different LTE-U modes.]{

\begin{minipage}[b]{0.5\textwidth}

\includegraphics[width=10cm]{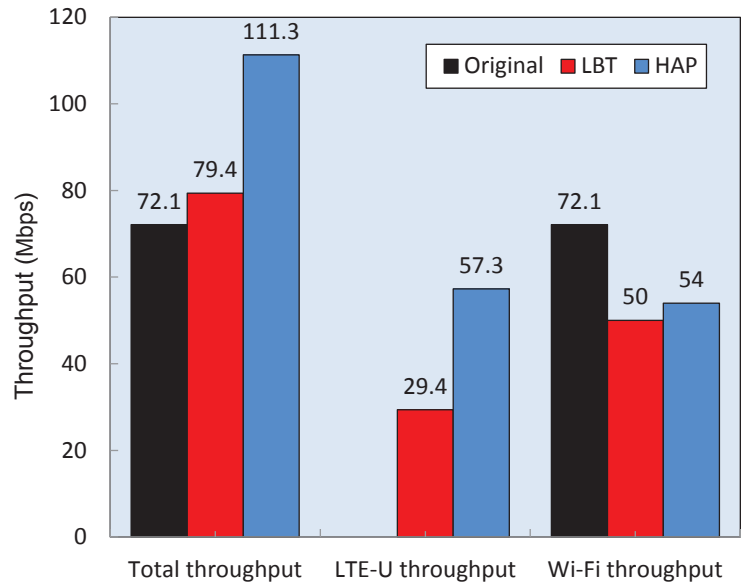}

\end{minipage}
}
\caption{Simulation results.}
\end{figure}


In Fig. 4(b), we further compare the
LTE-U throughput, the Wi-Fi throughput, and the overall system throughput under the three different network scenarios.
We randomly place $30$ Wi-Fi users and $10$ LTE-U users within the HAP coverage area.
In addition, the results in Fig. 4(b) illustrate that the proposed HAP architecture
achieves higher LTE-U throughput, higher Wi-Fi throughput, and higher overall system throughput when compared
with the LBT protocol.
Under HAP, the overall system throughput can be $50\%$ plus with only a mild loss of Wi-Fi throughput when compared against the original pure Wi-Fi network.
On the other hand, the LBT scheme not only exhibits a noticeable throughput  loss in Wi-Fi,  but also a significant drop in
the overall throughput than the proposed HAP network.
Clearly, our HAP serviced network can better utilize the unlicensed spectrum while better protecting the Wi-Fi user experience.

Furthermore, the improvement of unlicensed LTE-U throughput is also significant compared with the standalone licensed LTE throughput. Suppose the licensed bandwidth in each HAP is the same as the unlicensed bandwidth, i.e., $20$ MHz. Then we assume that one LTE user's rate within a typical 4G-LTE connection is at $8$ Mbps, which is within the average per-user throughput in practical LTE systems from $3$ to $10$ Mbps.
By allowing this user to be one of the $10$ users to access LTE-U, its throughput can be improved by as much as $71.6\%$ in the proposed HAP-based LTE-U network, rather than by $36.8\%$ in the LBT-based LTE-U network. We also test the case with $40$ Wi-Fi users, in which the proposed HAP-based LTE-U network still achieves as much as $57.3\%$ throughput gain, better than $27.8\%$ in the LBT-based LTE-U network.

Overall, we demonstrate that the shared spectrum access by LTE-U and Wi-Fi is more efficient than the conventional LBT protocol because of the more efficient LTE-U channel access. Furthermore, our proposed LTE-U embedding protocols are easy to implement.
They can be directly integrated into existing IEEE 802.11 networks without any Wi-Fi adjustment.  In fact, non-LTE users can
function in normal mode and appear fully oblivious to the presence of any LTE-U users.  Our test results
of the proposed LTE-U access protocol establish its network performance advantages in terms of sum throughput and per-user Wi-Fi throughput.

\section{Further Works and Open Research Issues}

Although the proposed HAP frameworks can substantially improve the spectrum efficiency and facilitate the deployment of LTE-U,
there still exist several open issues worth investigating:

\noindent
\textbf{Interference management}:
The first issue is the interference management in the unlicensed spectrum. Although we require HAPs to scan for clear
unlicensed band for LTE-U embedding,  the criteria for such decision deserve further detailed study, particularly
for dense Wi-Fi deployment areas.  Without a truly clean and open unlicensed channel, LTE-U has to reuse certain channels and
manage interference even during CFP.  In terms of interoperability, HAPs belonging to
different operators may compete for the same unlicensed bands and, if unregulated, could lead to major
failure of LTE-U. Therefore, problems involving channel selection and interoperability must be investigated.

\noindent
\textbf{CFP and CP Allocation}:
Another question lies in the resource division between CFP and CP.  For example,
when many hybrid users are admitted into the LTE-U system from Wi-Fi, it is natural to allocate longer CFP.
However, it may substantially degrade the performance of remaining Wi-Fi users relying on the shorter CP.
Conversely, if insufficient CFP is allocated,  per-user LTE-U throughput would suffer whereas precious resources
on the Wi-Fi side may be under-utilized.  It is therefore important to optimize the CFP and CP parameters (e.g. period, duration)
according to the network environment. The optimization depends not only on the number of users but also
on co-channel interference from adjacent cells or networks as well as the user traffic load.

\noindent
\textbf{Multiple HAP cooperation and clustering}:
Multiple HAPs 
may be deployed within the coverage of a macro-cell. In this case, how to cluster and jointly
control HAPs for better system performance and how to allocate wireless resources among
different HAPs are important open problems. Distributed resource optimizations based on
game theoretic models and machine learning approaches are expected to provide important insight.


\section{Conclusion}
This article aims to propose novel frameworks for the deployment of LTE-U in the
unlicensed spectrum. We first reviewed various operation modes and coexistence techniques for LTE and Wi-Fi
networks, and then presented a novel concept of HAP by combining SBS and Wi-Fi AP functionality to
overcome many thorny issues in LTE-U and Wi-Fi coexistence.
Our focus for embedding LTE-U fully within the present Wi-Fi protocol is motivated by the natural need
to avoid retrofitting Wi-Fi networks to accommodate LTE-U channel sharing.
We as well proposed two basic LTE-U embedding approaches within the existing Wi-Fi MAC protocol based on the novel HAP framework.
Our simulation results demonstrated the substantial throughput advantage of the proposed LTE-U embedding mechanisms
over existing methods.
We further identified several open research issues for fine-tuning and optimizing our proposed LTE-U embedding
mechanisms in practical implementation.


\end{document}